\documentclass[aps,prd,preprint,nofootinbib,superscriptaddress,tightenlines]{revtex4}
\usepackage{amsfonts}
\usepackage{mathrsfs}
\usepackage{graphicx}
\usepackage{amsmath}
\usepackage{amssymb}
\usepackage{multirow}
\usepackage{subfigure}
\usepackage{epsfig}
\usepackage{graphicx}
\usepackage{booktabs}
\usepackage{array}
\usepackage{tabularx}
\usepackage{slashed}
\usepackage{ulem}

\newcommand{\bqa}{\begin{eqnarray}}
\newcommand{\eqa}{\end{eqnarray}}
\newcommand{\beq}{\begin{equation}}
\newcommand{\eeq}{\end{equation}}
\graphicspath{{fig/}{dia/}} \DeclareGraphicsExtensions{.eps}
\begin{document}
\baselineskip 20pt
\title{Pseudoscalar heavy quarkonium production\\ in heavy ion ultraperipheral collision}

\author{Jun Jiang}
\email{jiangjun87@sdu.edu.cn}
\affiliation{School of Physics, Shandong University, Jinan, Shandong 250100, China}

\author{Shi-Yuan Li}
\email{lishy@sdu.edu.cn}
\affiliation{School of Physics, Shandong University, Jinan, Shandong 250100, China}

\author{Xiao Liang}
\email{202120290@mail.sdu.edu.cn}
\affiliation{School of Physics, Shandong University, Jinan, Shandong 250100, China}

\author{Yan-Rui Liu}
\email{yrliu@sdu.edu.cn}
\affiliation{School of Physics, Shandong University, Jinan, Shandong 250100, China}

\author{Cong-Feng Qiao}
\email{qiaocf@ucas.ac.cn}
\affiliation{School of Physical Sciences, University of Chinese Academy of Sciences, YuQuan Road 19A, Beijing 100049, China}
\affiliation{CAS Key Laboratory of Vacuum Physics, Beijing 100049, China}

\author{Zong-Guo Si}
\email{zgsi@sdu.edu.cn}
\affiliation{School of Physics, Shandong University, Jinan, Shandong 250100, China}

\author{Hao Yang}
\email{yanghao2023@scu.edu.cn}
\affiliation{College of Physics, Sichuan University, Chengdu, Sichuan 610065, China}

\begin{abstract}

The inclusive production of pseudoscalar heavy quarkoniua ($\eta_c,~\eta_b$ and $B_c$) via photon-photon fusion in heavy ion ultraperipheral collision (UPC) are calculated to QCD next-to-leading order in the framework of non-relativistic QCD (NRQCD). 
The total cross section of $\eta_c$ produced in Pb-Pb UPC is 194 $\mathrm{nb}$ and 1275 $\mathrm{nb}$ at nucleon-nucleon c.m. energies $\sqrt{S_{\mathrm{NN}}}=$ 5.52 TeV and 39.4 TeV, respectively.
The cross sections for $\eta_b$ and $B_c$ mesons are more than two to three orders of magnitude smaller.
We make a detailed phenomenological analysis on the $\eta_c$ production; 
the uncertainties caused by the renormalization scale and the charm quark mass, the cross sections in other ultraperipheral nucleon-nucleon colliding systems, and the transverse momentum distribution are discussed.
At the coming HL-LHC and future FCC, the heavy ion UPC opens another door of the study on the production of heavy quarkonium.

\vspace {2mm} 
\noindent {Keywords: Heavy quarkonium, QCD next-to-leading order, nucleus-nucleus ultraperipheral collisions}
\end{abstract}

\maketitle

\section{INTRODUCTION}
\label{sec:introduction}

The heavy quarkonium is a bound state which consists of a heavy quark and a heavy antiquark, i.e. the $Q\bar{Q}$ ($Q=c,\, b$ quarks) bounding systems.
The production of heavy quarkonium consists of the perturbative production of the heavy quarks pair $Q\bar{Q}$ and its nonperturbative hadronization into the heavy quarkonium $\mathcal{H}(Q\bar{Q})$. 
The non-relativistic QCD (NRQCD) factorization formalism \cite{Bodwin:1994jh,Petrelli:1997ge} can be used to study the quarkonium production and decay, where the perturbative and nonperturbative effects are factored out. 
The creation of the binding $Q\bar{Q}$ intermediate state with definite $J^{PC}$ and color configuration can be calculated perturbatively under the double series of the strong coupling $\alpha_s$ and the relative velocity $v$ between the heavy quark $Q$ and antiquark $\bar{Q}$.
While the nonperturbative hadronization of the intermediate states into the physical heavy quarkonium $\mathcal{H}(Q\bar{Q})$ is described by the long-distance matrix elements, which can be extracted from experiments, or calculated in lattice QCD \cite{Bodwin:1996tg}, or related to potential models \cite{Richardson:1978bt,Buchmuller:1980bm,Eichten:1994gt}. 

In heavy ion collisions, due to the large proton charge number $Z$, the highly relativistic ions become a strong source of electromagnetic radiation, which can be considered as fluxes of quasi-real photons in the equivalent photon approximation \cite{vonWeizsacker:1934nji,Williams:1934ad}.
Thus, the heavy ion collisions, say Pb-Pb collision, can be used to study photon-proton, photon-Pb and photon-photon interactions in the ultraperipheral collisions (UPC) \cite{Baur:2001jj,Bertulani:2005ru,Baltz:2007kq}.
In this article, we focus on the photon-photon fusion in Pb-Pb UPC. 
In comparison with the central heavy ion collision, the quasi-real photon interaction in UPC has low event multiplicity because the ion impact parameter is larger than twice the ion radius and the ions are kept unbroken, where the interaction is pure electromagnetic.
In comparison with the photon-photon fusion in proton-proton collision, the photon density is in particular enhanced by the squared ion charge $Z^2$ for each ion, leading to an overall $Z^4$
enhancement in the production rate.
In particular, the photon-initiated production in UPC has no noisy hadron background and has no complicated dependence on the parton distribution fucntions in the hardronic producion \cite{Baltz:2007kq,Contreras:2015dqa}.
In comparison with the photon-photon interaction at an electron-positron collider, one could expect that the de-excitation photons with energies up to 80 GeV and 600 GeV \cite{Shao:2022cly} can be emitted in Pb-Pb UPC at the high luminosity Large Hadron Collider (HL-LHC) \cite{Bruce:2018yzs,Klein:2020nvu,dEnterria:2022sut} and the Future Circular Collider (FCC) \cite{FCC:2018vvp,Dainese:2016gch}, respectively.
Therefore, the photon-initiated UPC processes should be another good laboratory to study the production of heavy quarkonium.

The production of heavy quarkonium in UPC has rich physics.
The quarkonium photonproduction off the proton in p–Pb UPC provides a direct tool to study the gluon density at low Bjorken-x, and quarkonium photonproduction in Pb–Pb UPC can be used to probe the nuclear gluon shadowing effects.
Different experiments at LHC conduct various measurements on the photonproduction of vector heavy quarkonium, $J/\psi,\,\psi(2S),\,\Upsilon$, in the UPC processes \cite{Kryshen:2014eha,Andronic:2015wma,Przybycien:2018tza}.
There are also several theoretical predictions on the photoproduction of heavy quarkonium, see Refs. \cite{Guzey:2016piu,Andrade:2022rbn} as examples.
For the processes by photon-photon fusion in Pb-Pb UPC, ATLAS \cite{ATLAS:2017fur} and CMS \cite{CMS:2018qbh} obtain an evidence for light-by-light scattering, and the $\gamma\gamma \longrightarrow l^+l^-$ ($l=e,\,\mu$) processes is explored by ALICE \cite{ALICE:2013wjo} and ATLAS \cite{ATLAS:2016vdy}.
In addition, the effects of linear polarization of photon and impact parameter dependence of the azimuthal asymmetry for $\gamma\gamma \longrightarrow l^+l^-$ in Pb-Pb UPC are studied \cite{Li:2019yzy,Li:2019sin}.
Up to date, there are no experements on the heavy quarkonium production via photon-photon fusion in UPC, while theoretical study takes its steps.
The  production of double $J/\psi$ via photon-photon fusion is calculated in Refs. \cite{Qiao:2001wv,Baranov:2012vu,Goncalves:2015sfy}, and the production of $S$ and $P$ wave $B_c$ mesons via photon-photon fusion is calculated in Refs. \cite{Berezhnoy:1994bb,Berezhnoy:1995ay}.
And the $\eta_c$ production in photon-photon interaction in p-p collision ($\sqrt{s}=13$ TeV) and p-Pb ($\sqrt{s}=18.1$ TeV) UPC is studied at the leading order in $\alpha_s$ \cite{Goncalves:2018yxc}.
In our previous works \cite{Yang:2022yxb}, the production of pseudoscalar heavy quarkonia ($\eta_c$, $\eta_b$ and $B_c$) via photon-photon interaction at an $e^{+}e^{-}$ collider are calculated up to the next-to-leading order (NLO) in $\alpha_s$ within NRQCD framework. 
In this manuscript, we study the production of pseudoscalar heavy quarkonia via photon-photon interaction in Pb-Pb UPC at NLO in $\alpha_s$. The advantages in UPC would shed lights on the study of the physics in heavy quarkonium production.

The rest of the paper is organized as follows. 
In section \ref{sec:formulation}, we introduce the calculation formalism for the inclusive production of heavy quarkonia in UPC at the NLO accuracy within the NRQCD framework. 
In section \ref{sec:data}, the numerical results, their uncertainties, and the differential distribution are discussed. 
Sec.\ref{sec:summary} is reserved for a summary.

\section{FORMULATION}
\label{sec:formulation}

In the equivalent photon approximation (EPA) formulism 
\cite{vonWeizsacker:1934nji,Williams:1934ad}, the cross section for the final state $X$ via photon fusion in an ultraperipheral collision (UPC) of heavy ions $\mathrm{A}$ and $\mathrm{B}$, $\sigma(\mathrm{A} \mathrm{B} \stackrel{\gamma \gamma}{\longrightarrow} \mathrm{A} \mathrm{B} X)$, can be factorized into a convolution of the equivalent photon spectra and the cross section for $X$ production in photon fusion $\hat{\sigma}({\gamma\gamma \longrightarrow X})$,
\begin{equation}
\sigma(\mathrm{A} \mathrm{B} \stackrel{\gamma \gamma}{\longrightarrow} \mathrm{A} \mathrm{B} X)=\int \frac{d x_1}{x_1} \frac{d x_2}{x_2} f(x_1) f(x_2)   \times \mathrm{d} \hat{\sigma}({\gamma \gamma \rightarrow X}),
\label{eq:totalXS}
\end{equation}
where $x_{i} = E_i/E_{\mathrm{beam}}$ is the ratio of the photon energy $E_i$ to the beam energy $E_{\mathrm{beam}}$. 
For an UPC, the spectrum of an equivalent photon $f(x)$ is a function of the impact parameter $b$ \cite{Cahn:1990jk,Baur:1990fx}. 
Integrating over the impact parameter $b$ from $b_{\mathrm{min}}$ to infinity, the photon spectrum $f(x)$ produced by an ion of charge $Z$ reads \cite{Jackson:1999,Shao:2022cly}
\begin{equation}
f(x)=\frac{2 \alpha Z^2}{\pi}\left[\chi K_0(\chi) K_1(\chi)-\left(1-\gamma_{\mathrm{L}}^{-2}\right) \frac{\chi^2}{2}\left(K_1^2(\chi)-K_0^2(\chi)\right)\right],
\label{eq:photonpdf}
\end{equation}
where the variable $x$ is absorbed into $\chi \equiv x m_N b_{\mathrm{min}}$, in which $m_N$ is the nucleon mass and the minimum of impact parameter $b_{\mathrm{min}}$ will be set as the nuclear radius. 
And $K_0,~K_1$ are the modified Bessel functions of the second kind of zero and first order, respectively. $\gamma_L=E_{\mathrm{beam}}/m_N$ is the Lorentz factor. 

It is worth noting that the heavy ions $\mathrm{A}$ and $\mathrm{B}$ may dissociate due to the secondary soft hadronic interaction. 
In our formulation, we have set the heavy ions to always remain intact, i.e. the survival probability of heavy ions is 100\%. 
Under this assumption, the two initial photon distribution just factorizes as two parton-distribution-function-like spectrum $f(x)$, otherwise a probability to have no inelastic hadronic interaction as a function of compact parameter $b$ has to be embedding into Eq. \eqref{eq:totalXS} \cite{Shao:2022cly}. 
Thus, our present estimations for cross sections could be treated as an upper limit.
An explicit example of this dissociation effect in Ref. \cite{Knapen:2016moh} shows that the correction is less than 20\%.

In the calculation, it's more convenient to display Eq. \eqref{eq:totalXS} in the variables of a ratio $x$ and the center-of-mass (c.m.) energy of two photons $W_{\gamma\gamma}$,
\begin{equation}
    \sigma(\mathrm{A} \mathrm{B} \stackrel{\gamma \gamma}{\longrightarrow} \mathrm{A} \mathrm{B} X)= 2 \int \frac{\mathrm{d} x}{x} \frac{\mathrm{d} W_{\gamma\gamma}}{W_{\gamma\gamma}} f(x) f(\frac{W_{\gamma\gamma}^2}{x S_{\mathrm{NN}}})   \times \mathrm{d}\hat{\sigma}({\gamma \gamma \rightarrow X}),
\end{equation}
where $x$ can be either $x_1$ or $x_2$ integrating from ${W_{\gamma\gamma}^2}/{S_{\mathrm{NN}}}$ to 1 with $\sqrt{S_{\mathrm{NN}}}=2E_{\mathrm{beam}}$ being the nucleon-nucleon c.m. energy. 
And the $W_{\gamma\gamma}$ runs from the sum of masses of particles in the final states to a maximum $W_{\gamma\gamma}^{\mathrm{max}}$ which we take the same values as the program gamma-UPC \cite{Shao:2022cly}.

In this paper, we consider the inclusive production of pseudoscalar heavy quarkoniua via photon-photon fusion in UPC, $\gamma+\gamma\to \mathcal{H}(Q_1\bar{Q}_2)+Q_2+\bar{Q}_1$, where $\mathcal{H}(Q_1\bar{Q}_2) = \eta_c$, $\eta_b$ or $B_c$ and $Q_{1,2}$ are the heavy charm or bottom quarks accordingly. 
Up to the leading order (LO) in QED and next-to-leading order (NLO) in QCD, the differential cross section consists of three parts,
\begin{equation}
    \mathrm{d}\hat{\sigma}(\gamma \gamma \longrightarrow \mathcal{H}(Q_1\bar{Q}_2) + \bar{Q}_1 + Q_2) = \mathrm{d}\hat{\sigma}_{\mathrm{born} } + \mathrm{d}\hat{\sigma}_{\mathrm{virtual}} + \mathrm{d}\hat{\sigma}_{\mathrm{real}}.
    \label{eq:subprocessXS}
\end{equation}
In the c.m. frame of two photons, the differential cross section at born level, its QCD virtual correction and QCD real correction take the forms,
\begin{equation}
    \begin{split}
        &\mathrm{d}\hat{\sigma}_{\mathrm{born}}=\frac{1}{2 W_{\gamma\gamma}}\overline{\sum}|\mathcal{M}_{\mathrm{tree}}|^{2} \mathrm{d}{\mathrm{PS}}_{3}\ ,\\
        &\mathrm{d}\hat{\sigma}_{\mathrm{virtual}}=\frac{1}{2 W_{\gamma\gamma}}\overline{\sum}2{\mathrm{Re}}(\mathcal{M}^{*}_{\mathrm{tree}}\mathcal{M}_{\mathrm{oneloop}}) \mathrm{d}{\mathrm{PS}}_{3}\ ,\\
        &\mathrm{d}\hat{\sigma}_{\mathrm{real}}=\frac{1}{2 W_{\gamma\gamma}}\overline{\sum}|\mathcal{M}_{\mathrm{real}}|^{2} \mathrm{d}{\mathrm{PS}}_{4}\ ,
    \end{split}
    \label{eq:XSinthree}
\end{equation}
where $\overline{\sum}$ takes the sum over the polarizations and colors of final states and average over spins of initial states, $\mathrm{d}{\mathrm{PS}}_{n}$ denotes the differential $n$-body phase space of the final states.

The Feynman diagrams and their amplitudes at the parton level for two heavy quark pairs produced by photon-photon fusion can be generated by FeynArts \cite{Hahn:2000kx}. At the leading order of relative velocity expansion, the projection for a heavy quark pair $(Q_1\bar{Q}_2)$ into a spin-singlet and color-singlet quarkonium $\mathcal{H}(Q_1\bar{Q}_2)$ can be performed by the simple replacement \cite{Bodwin:1994jh,Petrelli:1997ge},
\begin{equation}
       v_{\bar{Q}_2} \bar{u}_{Q_1} \longrightarrow \frac{\psi(0)}{2\sqrt{m_{\mathcal{H}}}} \gamma^5 (\slashed{p}_{\mathcal{H}} + m_{\mathcal{H}})\otimes \frac{\delta_{ij}}{\sqrt{N_{c}}},
       \label{eq:projection}
\end{equation}
where $m_{\mathcal{H}}$ and $p_{\mathcal{H}}$ are the mass and momentum of the quarkonium respectively, $\delta_{ij}$ guarantees the color-singlet of the heavy quarkonium, $N_c=3$ is the quark color number. And $\psi(0)$ is the Schr$\mathrm{\ddot{o}}$dinger wave function at the origin, which can be related to its radial wave function at the origin $R(0)$, $\psi(0)=R(0)/ \sqrt{4 \pi}$.

The analytical calculation of the virtual and real corrections to the differential cross sections in Eq. (\ref{eq:XSinthree}) is non-trivial.
We use the dimensional regularization to regularize the ultraviolet (UV) divergences in the virtual correction, which behaves as $1/\epsilon_{\mathrm{UV}}$. 
The UV divergences can be canceled by the corresponding counter terms. It's worth noting that counter terms related to the gluon field $Z_3$ will cancel each other out because there is no Feynman diagrams with gluon external leg in our case.
To isolate the IR divergences in  virtual correction, we use the method proposed in Ref. \cite{Kramer:1995nb}.
We introduce an infinite small mass $\lambda$ for the gluon to regularize the infrared (IR) divergences in the virtual correction, which has the form of $\mathrm{log}(\lambda^2/m^2)$ with $m$ being the heavy quark mass.\footnote{In dimension regularization scheme, the IR divergence behaves as $1/\epsilon_{\mathrm{IR}}$ instead.} 
Note that among the counter terms only the one related to the heavy quark field $Z_2$ has IR singularity.
There are Coulomb singularities in the virtual correction when a gluon is exchanged between the constituent heavy quark pair which projects into the heavy quarkonium. It can also be regularized by the infinitesimal gluon mass $\lambda$, and the Coulomb singularity is proportional to the squared born amplitudes,
\begin{equation}
    2{\mathrm{Re}}(\mathcal{M}^{*}_{\mathrm{tree}}\mathcal{M}_{\mathrm{oneloop}}) \overset{\mathrm{Coulomb}}{\sim} |\mathcal{M}_{\mathrm{tree}}|^{2} \frac{2\alpha_s C_F m}{\lambda},
    \label{eq:coulomb}
\end{equation}
which should be absorbed into the wave function of the heavy quarkonium. In the case of $B_c$ production, the $m$ in Eq. (\ref{eq:coulomb}) should be replaced by the reduced mass $\frac{2m_bm_c}{m_b+m_c}$. 
There are IR divergences in the real correction, and we also introduce the infinitesimal gluon mass $\lambda$ to regularize them. To isolate the IR singularities, the dipole subtraction method formulated in Ref. \cite{Dittmaier:1999mb} is adopted.
We introduce an auxiliary subtraction $|\mathcal{M}_{\mathrm{sub}}|^{2}$ which holds the same asymptotic behavior as the real squared amplitude $|\mathcal{M}_{\mathrm{real}}|^{2}$ in the soft limit.
Then cross section of the modified real correction reads
\begin{equation}
    \hat{\sigma}_{\mathrm{real}}^- = \frac{1}{2\hat{s}}\int d{\mathrm{PS}}_{4} (\overline{\sum}|\mathcal{M}_{\mathrm{real}}|^{2} - |\mathcal{M}_{\mathrm{sub}}|^{2}),
    \label{eq:modifiedreal}
\end{equation}
which has no IR singularities at each point of phase space and can be evaluated by setting $\lambda=0$.
The contribution of the subtraction term to the real correction reads 
\begin{equation}
    \hat{\sigma}_{\mathrm{real}}^+ = \frac{1}{2\hat{s}}\int d{\mathrm{PS}}_{4} |\mathcal{M}_{\mathrm{sub}}|^{2} = \frac{1}{2\hat{s}}\int d{\mathrm{PS}}_{3} \int [dp_g]|\mathcal{M}_{\mathrm{sub}}|^{2}.
    \label{eq:subtractionXS}
\end{equation}
With an appropriate construction of the subtraction term, the integral over the emitting gluon $ \int [dp_g]|\mathcal{M}_{\mathrm{sub}}|^{2}$ can be carried out analytically, and the remaining integral becomes a 3-body phase space integration like the virtual correction. 
The IR divergences from $ \int [dp_g]|\mathcal{M}_{\mathrm{sub}}|^{2}$ shall cancel those in the virtual amplitudes $2{\mathrm{Re}}(\mathcal{M}^{*}_{\mathrm{tree}}\mathcal{M}_{\mathrm{oneloop}})$, which leads to the IR-free result at NLO.
The total contribution from the real correction in this method is the sum
\begin{equation}
    \hat{\sigma}_{\mathrm{real}}=\hat{\sigma}_{\mathrm{real}}^- + \hat{\sigma}_{\mathrm{real}}^+.
    \label{eq:realXS}
 \end{equation}
For more details of the analytical calculation of the differential cross section at LO and at NLO, we suggest that the readers refer to our previous work \cite{Yang:2022yxb}. In this paper, we focus on the phenomenological analysis on the production of pseudoscalar heavy quarkoniua in the heavy ion ultraperipheral collision.

\section{NUMERICAL RESULTS}
\label{sec:data}

To obtain the numerical results, we first determine the input parameters.
We have parameters related to the UPC of heavy-ions, taking the Pb-Pb collision at $\sqrt{S_{\mathrm{NN}}}=5.52$ TeV as an example, 
\begin{align}
W_{\gamma\gamma}^{\mathrm{max}}=160\ {\rm GeV},\quad Z=82,\quad m_N=0.9315\ {\rm GeV},\quad b_{min}=7.1\ {\rm fm}.
\end{align}
For the input parameters in cases of other heavy-ion collisons  at other nucleon-nucleon c.m. energies, we take the same values as the program gamma-UPC \cite{Shao:2022cly}.
We also have parameters related to the NLO calculation of the production of heavy quarkonium via photon-photon fusion,
\begin{align}
&\alpha=1/137.065,\quad m_c=1.5\ {\rm MeV},\quad m_b=4.8\ {\rm GeV},\nonumber \\
&|R_{\eta_c}^{\rm LO}(0)|^2=0.528\ {\rm GeV}^3,\quad |R_{\eta_c}^{\rm NLO}(0)|^2=0.907\ {\rm GeV}^3; \nonumber \\
&|R_{\eta_b}^{\rm LO}(0)|^2=5.22\ {\rm GeV}^3,\quad |R_{\eta_b}^{\rm NLO}(0)|^2=7.48\ {\rm GeV}^3; \nonumber \\
&|R_{B_c}(0)|^2=1.642\ {\rm GeV}^3.
\end{align}
Here, the $R_{\eta_c}(0)$, $R_{\eta_b}(0)$ and $R_{B_c}(0)$ are the radial wave functions at the origin for the S-wave states $\eta_c$, $\eta_b$ and $B_c$, respectively. The values for $\eta_c$ and $\eta_b$ are extracted from their leptonic decays, while the one for $B_c$ is estimated using the Buchmueller-Tye potential model \cite{Eichten:1994gt}. 
Note that we have different values for radial wave functions at the origin in the LO and NLO estimations for $\eta_c$ and $\eta_b$. This is because we employ one-loop formula of running strong coupling constant $\alpha_s$ to extract the radial wave functions at the origin in the LO calculation, but two-loop formula in the NLO calculation \cite{Yang:2022yxb}. 
The running two-loop $\alpha_s(\mu)$ as a function of energy scale $\mu$ reads 
\begin{equation}
    \alpha_{s}(\mu)=\frac{4\pi}{\beta_{0}\ln{\mu^2/\Lambda^2_{\rm QCD}}}\left[1-\frac{\beta_{1}\ln{\ln{\mu^2/\Lambda^2_{\rm QCD}}}}{\beta_{0}^2\ln{\mu^2/\Lambda^2_{\rm QCD}}}\right],
\end{equation}
in which the QCD $\beta$-functions  $\beta_0=\tfrac{11}{3}C_A-\tfrac{4}{3}T_Fn_f$, and $\beta_{1}=\frac{34}{3}C^{2}_{A}-4C_{F}T_{F}n_{f}-\frac{20}{3}C_{A}T_{F}n_{f}$. We take $n_f=4$, $\Lambda_{\rm QCD}=297$ MeV for $\eta_c$ production, and $n_f=5$, $\Lambda_{\rm QCD}=214$ MeV for $\eta_b$ and $B_c$ cases.

\begin{table}[ht]
    \caption{The LO and NLO total cross sections (in nb) for $\eta_c+c+\bar{c}$, $\eta_b +b+\bar{b}$ and $B_c +b+\bar{c}$ production via photon-photon fusion in ultraperipheral Pb-Pb collision at $\sqrt{S_{\mathrm{NN}}}=$ 5.52 and 39.4 TeV. Here,  scale $\mu=\sqrt{m_H^2 + p_{t}^2}$ with $m_H$ being the mass of heavy quarkonia, and the transverse momentum cut $ 1\ {\rm GeV} \le p_{t} \le 50$ GeV is employed.}
    \begin{center}
       \begin{tabular}{p{5cm}<{\centering} p{3cm}<{\centering} p{3cm}<{\centering} p{3cm}<{\centering}}
        \toprule
        \hline
            processes  & $\eta_c +c\bar{c}$   &  $\eta_b +b\bar{b}$  & $B_c + b\bar{c}$ \\
        \hline
        $\sigma_{\rm LO}$ ($5.52 $ TeV)&
        $1.7\times 10^2$ &
        $0.034$ &
        $0.57$ \\
        $\sigma_{\rm NLO}$ ($5.52 $ TeV)&
        $1.9\times 10^2$ &
        $0.027$ &
        0.47 \\
        K-factor &
        $1.2$ &
        $0.78$ &
        0.83 \\
         $\sigma_{\rm LO}$ ($39.4 $ TeV)&
        $1.1\times 10^3$ &
        $0.46$ &
        6.5 \\
         $\sigma_{\rm NLO}$ ($39.4 $ TeV)&
        $1.3\times 10^3$ &
        $0.30$ &
        5.8\\
         K-factor &
        1.2 &
        $0.66$ &
       0.90 \\
        \botrule
      \end{tabular}
    \end{center}
    \label{all process}
\end{table}

In table \ref{all process}, we present the total cross sections for the inclusive production of pseudoscalar quarkonia ($\eta_c,\, \eta_b,\, B_c$) via photon-photon fusion in Pb-Pb ultraperipheral collision. 
We consider two nucleon-nucleon c.m. energies $\sqrt{S_{\mathrm{NN}}}=$ 5.52 and 39.4 TeV, which are typical collision energies at current LHC and future HL-LHC and FCC.
The K-factor is defined as the ratio of cross sections of NLO to that of LO, $\sigma_{\mathrm{NLO}}/\sigma_{\mathrm{LO}}$. 
Numerical results indicate that the NLO corrections are significant.
The NLO corrections improve the cross sections of $\eta_c$ by about 16\% and 21\% for $\sqrt{S_{\mathrm{NN}}}=$ 5.52 and 39.4 TeV, respecively.
While the NLO corrections for both $\eta_b$ and $B_c$ are negative;
cross sections at NLO for $\eta_b$ decrease by 22\% and 43\% for $\sqrt{S_{\mathrm{NN}}}=$ 5.52 and 39.4 TeV respecively, and the two percentages are 17\% and 10\% for $B_c$.
Note that the vector $B_c^*$ signal can't be separated from pseudoscalar $B_c$ meson in experiment and will decay into the later electromagnetically with almost 100\% probability. We obtain the LO cross sections for $B_c^*$ meson as 7.0 $\mathrm{nb}$ and 65 $\mathrm{nb}$ for $\sqrt{S_{\mathrm{NN}}}=$ 5.52 TeV and 39.4 TeV respectively, which are about one order of magnitude greater than those of pseudoscalar $B_c$ accordingly.
Since the total cross section of the production of $\eta_c$ are much more than two orders of magnitude greater than those of other two processes, we focus on the phenomenological analysis of the $\eta_c$ production below.

\begin{table}[htb]
    \caption{The LO and NLO total cross sections (in nb) under typical renormalization scale values for $\eta_c+c+\bar{c}$ production via photon-photon fusion in ultraperipheral Pb-Pb collision at $\sqrt{S_{\mathrm{NN}}}=$ 5.52 and 39.4 TeV. Here, the transverse momentum cut $ 1\ {\rm GeV} \le p_{t} \le 50$ GeV is employed. And the scale $\mu=r\sqrt{4m_c^2 + p_{t}^2}$  varies by a factor of $r=\{0.5,1,2\}$.}
    \begin{center}
       \begin{tabular}{p{5cm}<{\centering} p{3cm}<{\centering} p{3cm}<{\centering} p{3cm}<{\centering}}
        \toprule
        \hline
            $r$  & $0.5$  &  $1$ & $2$ \\
        \hline
        $\sigma_{\rm LO}$ ($5.52 $ TeV)&
        $3.2\times 10^2$ &
        $1.7\times 10^2$ &
        $1.0\times 10^2$ \\
        $\sigma_{\rm NLO}$ ($5.52 $ TeV)&
        $3.2\times 10^2$ &
        $1.9\times 10^2$ &
        $1.3\times 10^2$ \\
         K-factor &
        $1.0$ &
        $1.2$ &
        $1.3$ \\
         $\sigma_{\rm LO}$ ($39.4 $ TeV)&
        $2.0\times 10^3$ &
        $1.1\times 10^3$ &
        $0.65\times 10^3$ \\
         $\sigma_{\rm NLO}$ ($39.4 $ TeV)&
        $2.1\times 10^3$ &
        $1.3\times 10^3$ &
        $0.86\times 10^3$ \\
         K-factor &
        $1.1$ &
        $1.2$ &
        $1.3$ \\
        \botrule
      \end{tabular}
    \end{center}
    \label{cccc r}
\end{table}

In table \ref{cccc r}, we show the uncertainties of the cross sections arising from the renormalization scale $\mu$ at both LO and NLO for $\eta_c+c+\bar{c}$ production. The scale $\mu=r\sqrt{4m_c^2 + p_{t}^2}$ varies by a factor of $r=\{0.5,1,2\}$.
It is found that the cross sections at LO and NLO decrease as the renormalization scale $\mu$ increases. 
More specially, the LO cross section at $\sqrt{S_{\mathrm{NN}}}=$ 5.52 TeV increases
by 91\% and decrease by 39\% for $\mu$ varies by a factor of 1/2 and 2, respectively.
At NLO, these two percentages becomes 65\% and 32\%. 
At $\sqrt{S_{\mathrm{NN}}}=$ 39.4 TeV, the LO cross section increases
by 90\% and decrease by 38\% for $\mu$ varies by a factor of 1/2 and 2, respectively.
While at NLO, these two percentages becomes 67\% and 32\%.
The NLO correction improve the $\mu$ dependence as we expect. 
We also find that the K-factors increase as scale $\mu$ increases, which indicates that the NLO corrections are more remarkable at high scale region.

In table \ref{cccc mc}, we present the uncertainties of the cross sections caused by the varying charm quark mass at both LO and NLO for $\eta_c+c+\bar{c}$ production.
Three charm quark masses $m_c=\{1.4,1.5,1.6\}$ GeV are adopted.
It is found that the cross sections at LO and NLO decrease as the charm quark mass increases. 
More specially, cross section at LO at $\sqrt{S_{\mathrm{NN}}}=$ 5.52 TeV increases
by 54\%  when $m_c$ decreases from 1.5 GeV to 1.4 GeV, and decreases by 34\% when $m_c$ grows from 1.5 GeV to 1.6 GeV. 
At NLO at $\sqrt{S_{\mathrm{NN}}}=$ 5.52 TeV, these two percentages becomes 51\% and 33\%.
At $\sqrt{S_{\mathrm{NN}}}=$ 39.4 TeV, cross section at LO increases by 50\% when $m_c$ decreases from 1.5 GeV to 1.4 GeV, and decreases by 32\% when $m_c$ grows from 1.5 GeV to 1.6 GeV.
While at NLO at $\sqrt{S_{\mathrm{NN}}}=$ 39.4 TeV, these two percentages becomes 46\% and 31\%.
For the K-factor, it grows slowly as the mass increases.

\begin{table}[ht]
    \caption{The LO and NLO total cross sections (in nb) under typical charm quark masses for $\eta_c+c+\bar{c}$ production via photon-photon fusion in ultraperipheral Pb-Pb collision at $\sqrt{S_{\mathrm{NN}}}=$ 5.52 and 39.4 TeV. Here, $\mu=\sqrt{4m_c^2 + p_{t}^2}$ and the transverse momentum cut $ 1\ {\rm GeV} \le p_{t} \le 50$ GeV is employed. And the charm quark masses are $m_c=\{1.4,1.5,1.6\}$ GeV.}
    \begin{center}
       \begin{tabular}{p{5cm}<{\centering} p{3cm}<{\centering} p{3cm}<{\centering} p{3cm}<{\centering}}
        \toprule
        \hline
            $m_c\ (\rm{GeV})$  & $1.4$  &  $1.5$ & $1.6$ \\
        \hline
        $\sigma_{\rm LO}$ ($5.52 $ TeV)&
        $2.6\times 10^2$ &
        $1.7\times 10^2$ &
        $1.1\times 10^2$ \\
        $\sigma_{\rm NLO}$ ($5.52 $ TeV)&
        $2.9\times 10^2$ &
        $1.9\times 10^2$ &
        $1.3\times 10^2$ \\
         K-factor &
        $1.1$ &
        $1.2$ &
        $1.2$ \\
         $\sigma_{\rm LO}$ ($39.4 $ TeV)&
        $1.6\times 10^3$ &
        $1.1\times 10^3$ &
        $0.72\times 10^3$ \\
         $\sigma_{\rm NLO}$ ($39.4 $ TeV)&
        $1.9\times 10^3$ &
       $1.3\times 10^3$ &
        $0.88\times 10^3$ \\
         K-factor &
        $1.2$ &
        $1.2$ &
        $1.2$ \\
        \botrule
      \end{tabular}
    \end{center}
    \label{cccc mc}
\end{table}

In table \ref{caca}, we show the LO and NLO total cross sections for $\eta_c+c+\bar{c}$ production via photon-photon fusion in various ultraperipheral nucleon-nucleon colliding systems.
Different colliding systems have different nucleon-nucleon c.m. energies: $\sqrt{S_{\mathrm{NN}}}=5.52,\,5.86,\,6.46,\,6.3,\,7.0$, and $7.0$ TeV for ions Pb, Xe, Kr, Ca, Ar, and O, respectively \cite{Shao:2022cly}.
Since the photon spectrum is proportional to the square of ion charge $Z^2$, the total cross section would be enhanced by $Z^4$ in an ultraperipheral nucleon-nucleon collision.
The ratios of the quartic ion charge $Z^4$ for ions  Pb, Xe, Kr, Ca, Ar, O are
\{$Z^4_{\mathrm{Pb}}/Z^4_{\mathrm{Xe}},\, Z^4_{\mathrm{Xe}}/Z^4_{\mathrm{Kr}}, \, Z^4_{\mathrm{Kr}}/Z^4_{\mathrm{Ca}},\, Z^4_{\mathrm{Ca}}/Z^4_{\mathrm{Ar}},\, Z^4_{\mathrm{Ar}}/Z^4_{\mathrm{O}}$ \}= \{5.3, 5.1, 11, 1.5, 26\}.
The corresponding ratios of cross sections at LO are \{$\sigma_{\mathrm{LO}}^{\mathrm{Pb}}/\sigma_{\mathrm{LO}}^{\mathrm{Xe}},\, \sigma_{\mathrm{LO}}^{\mathrm{Xe}}/\sigma_{\mathrm{LO}}^{\mathrm{Kr}}, \, \sigma_{\mathrm{LO}}^{\mathrm{Kr}}/\sigma_{\mathrm{LO}}^{\mathrm{Ca}},\, \sigma_{\mathrm{LO}}^{\mathrm{Ca}}/\sigma_{\mathrm{LO}}^{\mathrm{Ar}},\, \sigma_{\mathrm{LO}}^{\mathrm{Ar}}/\sigma_{\mathrm{LO}}^{\mathrm{O}}$ \}= \{4.1, 3.8, 7.7, 1.7, 18\}, while the corresponding ratios at NLO are \{4.2, 3.7, 7.8, 1.7, 18\}.
The ratios of the cross sections at LO and at NLO are approximately the same, and has the same evolution trend but smaller values as the ratios of quartic ion charge $Z^4$ for those ions.

\begin{table}[ht]
    \caption{The LO and NLO total cross sections (in nb) for $\eta_c+c+\bar{c}$ production via photon-photon fusion in various ultraperipheral nucleon-nucleon colliding systems. Here $\mu=\sqrt{4m_c^2 + p_{t}^2}$, and transverse momentum cut $ 1\ {\rm GeV} \le p_{t} \le 50$ GeV is employed for $\eta_c$. Note, different colliding systems have different nucleon-nucleon c.m. energies, see texts for details.}
    \begin{center}
       \begin{tabular}{p{2cm}<{\centering} p{2cm}<{\centering} p{2cm}<{\centering} p{2cm}<{\centering} p{2cm}<{\centering} p{2cm}<{\centering} p{2cm}<{\centering}} 
        \toprule
        \hline
             Nucleon &  Pb-Pb  &  Xe-Xe & Kr-Kr & Ca-Ca & Ar-Ar & O-O \\
        \hline
        $\sigma_{\rm LO}$ &
        $1.7\times 10^2$ &
        $41$ &
        $11$ &
        $1.4$ &
        $0.83$ &
        $0.046$\\
        $\sigma_{\rm NLO}$ &
        $1.9\times 10^2$ &
        $46$ &
        $12$ &
        $1.6$ &
        $0.94$ &
        $0.052$\\
         K-factor &
        $1.2$ & 
        $1.1$ &
        $1.1$ &
        $1.1$ &
        $1.1$ &
        $1.1$\\
        \botrule
      \end{tabular}
    \end{center}
    \label{caca}
\end{table}

\begin{figure}[!thbp]
    \centering
    \includegraphics[width=1\textwidth]{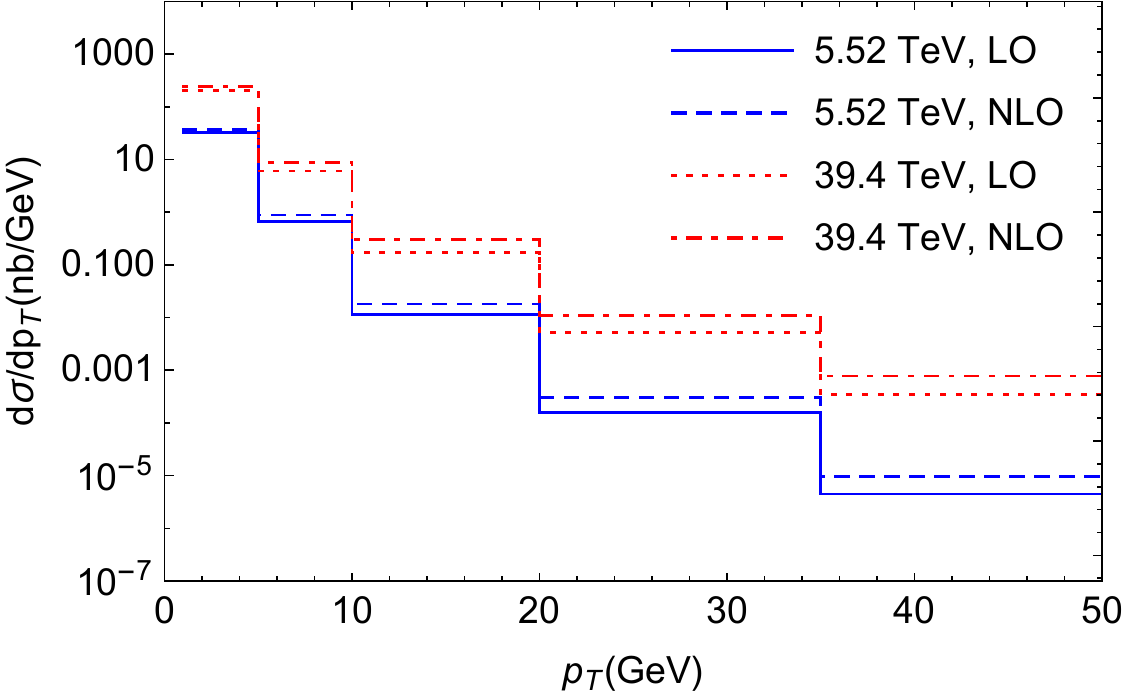}
    \caption{The differential transverse momentum distribution $d\sigma /dp_t$ of $\eta_c$ for the $\eta_c+c+\bar{c}$ production via photon-photon fusion in ultraperipheral Pb-Pb collision at $\sqrt{S_{\mathrm{NN}}}=$ 5.52 and 39.4 TeV. Here, $\mu=\sqrt{4m_c^2 + p_{t}^2}$ and the transverse momentum cut $ 1\ {\rm GeV} \le p_{t} \le 50$ GeV is employed.}
    \label{figptv}
\end{figure}

In figure \ref{figptv}, we present the differential distribution versus the transverse momentum of $\eta_c$ for $\eta_c+c+\bar{c}$ production via photon-photon fusion in ultraperipheral Pb-Pb collision.
The differential cross sections for both LO and NLO and at both $\sqrt{S_{\mathrm{NN}}}=$ 5.52 TeV and 39.4 TeV decrease monotonically as the transverse momentum $p_t$ of $\eta_c$ grows from 1 to 50 GeV. 
Additionally, the NLO corrections lead to an increasement in comparison with the LO results in the distributions, while their overall lineshapes are preserved.

We move to discuss the events for the inclusive production of pseudosclar quarkonia ($\eta_c,\, \eta_b,\, B_c$) via photon-photon fusion in Pb-Pb ultraperipheral collision.
Taking the integrated luminosity per typical run ${\cal L}_{\mathrm{int}}=5 \, {\mathrm{nb}}^{-1}$ and nucleon-nucleon c.m. energy $\sqrt{S_{\mathrm{NN}}}=5.52$ TeV at HL-LHC \cite{Shao:2022cly}, we roughly have $970^{+500}_{-320}$ events produced for $\eta_c$, where the uncertainties are caused by the charm quark mass only. And there are roughly 0 event for $\eta_b$ and 2 events for $B_c$ meson.
Taking the integrated luminosity per typical run ${\cal L}_{\mathrm{int}}=110 \, {\mathrm{nb}}^{-1}$ and nucleon-nucleon c.m. energy $\sqrt{S_{\mathrm{NN}}}=39.4$ TeV at FCC \cite{Shao:2022cly}, the events produced are roughly $(1.4^{+0.65}_{-0.43}) \times 10^5$, 30 and 640 events for $\eta_c, \, \eta_b$ and $B_c$ mesons, respectively.
One might observe the $\eta_c$ signals produced via photon-photon fusion in Pb-Pb UPC process in the comming HL-LHC experiment.

In the end, we discuss two subprocesses which might contribute to the inclusive cross sections of pseudoscalar quarkonium. There exists a direct subprocess $\gamma+\gamma\to\eta_c$ which is formally ahead of $\gamma+\gamma\to\eta_c+c+\bar{c}$ by two powers of $\alpha_{s}$ (and, similarly, $\gamma+\gamma\to\eta_b$). However, for the initial quasi-real photons, the $\eta_c$ produced by subprocess $\gamma+\gamma\to\eta_c$ has almost zero transverse momentum $p_{t}$. Thus, a small cut on $p_{t}$ would suppress the contribution from such subprocess. Alternatively, in the photon-induced UPC event which has clean QCD background, $\eta_c$ is the only final state for the exclusive subprocess $\gamma+\gamma\to\eta_c$, while for the inclusive subprocess $\gamma+\gamma\to\eta_c+c+\bar{c}$ it has extra quark jets. A trigger for extra jets in experiment will help to distinguish them.
There might also exist a contribution from nucleus-nucleus quasi-elastic scattering $A+A\to A+A+\gamma$ followed by a photon fragmentation $\gamma\to\eta_c+\gamma$. However, the rapidity coverage of such event should be spreaded near the colliding beams, while a photon-induced UPC event has large central rapidity coverage. Thus, an experiment cut on proper rapidity region would help to suppress the contribution from quasi-elastic nucleus-nucleus scattering.

\section{SUMMARY}
\label{sec:summary}

Within the NRQCD factorization formulism, we study the inclusive processes of $\eta_c +c+\bar{c}$, $\eta_b +b+\bar{b}$ and $B_c+b+\bar{c}$ through photon-photon fusion in ultraperipheral ion-ion collision up to QCD NLO. 
The results show that the NLO corrections are significant, which are presented in table \ref{all process}. 
We made a phenomenological analysis for $\eta_c +c+\bar{c}$ production in detail. 
The uncertainties of cross sections of $\eta_c$ caused by both the renormalization scale and the charm quark mass are shown in tables \ref{cccc r} and \ref{cccc mc}, respectively. We also explore the production of  $\eta_c$ at various ultraperipheral nucleon-nucleon colliding systems, which are presented in table \ref{caca}.
To make our analysis more useful to future experiments, the differential cross sections $d\sigma/dp_t$ versus transverse momentum of $\eta_c$ at $\sqrt{S_{\mathrm{NN}}}=$ 5.52 and 39.4 TeV are presented in figure \ref{figptv}.
The events for $\eta_c,\, \eta_b,\, B_c$ at future HL-LHC and FCC are also discussed, and it is feasible to observe $\eta_c$ at HL-LHC and future FCC.

The heavy ion UPC provides another good opportunity for the study of the production of heavy quarkonium. 
Our results show that sizable events for pseudoscalar $\eta_c$ can be produced in UPC process.
The events for vector $J/\psi$ would be even greater as a rule of thumb.
The absence of the background and uncertainty from QCD-initiated production would provide more efficient signal selection for the production of heavy quarkonium in UPC than in its hadronic production.
Meanwhile, the heavy ion UPC would provide much higher c.m. energies than the photon-photon interaction in electron-positron collision. 
Thus, the study of the production of heavy quarkonium in heavy ion UPC is worth anticipating at HL-LHC and future FCC.

\vspace{0.5cm} {\bf Acknowledgments}
This work is supported in part by the National Key Research and Development Program of China under Contracts No. 2020YFA0406400,
and in part by the National Natural Science Foundation of China (NSFC) under the Grants No. 12235008, No. 12321005 and No. 12275157.


\end{document}